\documentclass[11pt]{article}
\usepackage{amsmath}
\usepackage{amsfonts}
\usepackage{amssymb}
\usepackage{graphicx}

\def\bea{\begin{eqnarray}}
\def\eea{\end{eqnarray}}

\begin{document}
\begin{center}
\LARGE { \bf Logarithmic modes of  critical gravity in
de Sitter space-time
  }
\end{center}
\begin{center}
{\bf M. R. Setare\footnote{rezakord@ipm.ir} \\  H. Karimi}\\
 {Department of Science, Payame Noor University, Bijar, Iran }
 \\
 \end{center}
\vskip 3cm

\begin{abstract}
In this paper we consider the critical gravity in four dimensional
de Sitter space-time. We obtain logarithmic modes in the critical
point of the theory. Then we show that these logarithmic modes in de
Sitter space-time obey similar properties as the ones in
AdS-space-time. Our result in this paper indicate that critical
gravity theories in de Sitter space-times could lead to a de
Sitter/log CFT correspondence.

\end{abstract}

\newpage

\section{Introduction}
Holography is believed to be one of the fundamental principles of
the true quantum theory of gravity \cite{1,2}. An explicitly
calculable example of holography is the much studied Anti de Sitter
(AdS)/Conformal Field Theory (CFT) correspondence\cite{3}. We would
expect dS/CFT correspondence to proceed along the lines of Anti-de
Sitter /Conformal Field Theory (AdS/CFT) correspondence because de
Sitter spacetime can be obtained from anti-de Sitter spacetime by
analytically continuing the cosmological constant to imaginary
values. However, local and global properties of dS spacetime lead to
unexpected obstructions. Unlike AdS, the boundary of dS spacetime is
spacelike and its dual CFT is Euclidean. Moreover, dS spacetime does
not admit a global timelike Kiling vector. The time dependence of
the spacetime metric precludes a consistent definition of energy and
the use of Cardy formula to compute dS entropy. Finally dS/CFT
duality leads to boundary operators with complex conformal weights,
i.e. to a non-unitary CFT. In spite of these difficulties, some
progress towards a consistent definition of dS/CFT correspondence
has been achieved \cite{4}( see also \cite{5,6,7,8,9,10,11,12}).\\
It is well known that Einstein gravity suffers from the problem that
the theory is nonrenormalizable in four and higher dimensions.
Adding higher derivative terms such as Ricci and scalar curvature
squared terms makes the theory renormalizable at the cost of the
loss of unitarity \cite{13}. Usually the theories including the
terms given by the square of the curvatures have the massive spin 2
mode and the massive scalar mode in addition to the massless
graviton. Also the theory has ghosts due to negative energy
excitations of the massive tensor. In a special case that the
curvature squared term is given by the square of the Weyl tensor,
however, the massive scalar mode does not appear, which can be shown
by choosing an appropriate gauge condition. Recently,
quadratic-curvature actions with cosmological constant have been
introduced in four \cite{14} and $D$ \cite{15} dimensions. It was
found that there exist a choice of parameters for which these
theories possess one AdS background on which neither massive fields,
nor massless scalars propagate. By this special choice of the
parameters, which is called as a critical point, there appears a
mode which behaves as a logarithmic function of the distance.
Logarithmic modes in the framework of the higher-dimensional
critical gravity models were recently found in \cite{16, 17}. After
that Bergshoeff et. al \cite{18} shown that logarithmic modes are of
two types, spin 2 and Proca log modes. The number of independent
spin 2 log modes is given by the number of polarization states of a
massless spin 2 field, while the number of independent Proca log
modes is given by the number of polarization states of a massive
spin 1 field. They have obtained the logarithmic solutions of the
linearized $4-$dimensional critical gravity. They have considered
the linearization of the theory around anti de Sitter space-time.
The authors of \cite{18} have not considered the de Sitter case in
their work, so it would be interesting to find logarithmic modes,
when one consider the linearization of the theory around  de Sitter
space-time. In fact this is our job in this paper. We solve the
equation of motion in $4-$dimensional de Sitter background, and show
that the logarithmic modes appear as well as anti de Sitter
background. Then it would be interesting to see whether the log
modes in de Sitter space-time obey similar properties as the ones in
AdS-space-time, namely whether the properties of Eq.(64) and
especially Eq.(65) in \cite{18} still hold.  In the AdS-case, the
generator $H_1$ can be identified with the conformal energy operator
in the dual field theory. The property of Eq.(65) then indicates
that this conformal energy is not diagonalizable, which is the
defining property of a logarithmic CFT. Since the generator $H_1$ in
dS case is still correspond to the field theory Hamiltonian, by
finding an equation similar to the Eq.(65) in \cite{18}, we present
a strong indication that critical gravity theories in de Sitter
space-times could also lead to a de Sitter/log CFT correspondence.

\section{Logarithmic modes of critical gravity in de Sitter space-time}
We consider the following 4-dimensional gravity action
\begin{eqnarray}\label{1'}
S=\frac{1}{2\kappa^{2}}\int\sqrt{-g}d^{4}x
[R+\frac{6}{L^{2}}-\frac{1}{2m^2}(
R_{\mu\nu}R^{\mu\nu}-\frac{1}{3}R^2)-\frac{1}{4m^2} E_{GB}],
\end{eqnarray}
where $\kappa^{2}=8\pi G$, and
\begin{eqnarray}\label{2'}
E_{GB}=R^2-4R_{\mu\nu}R^{\mu\nu}+R_{\mu\nu\rho\sigma}R^{\mu\nu\rho\sigma},
\end{eqnarray}
is the the Gauss-Bonnet term. This term is the total derivative in
four dimensions. The action (\ref{1'}) can be rewritten as

\begin{eqnarray}\label{3'}
S=\frac{1}{2\kappa^{2}}\int\sqrt{-g}d^{4}x
[R+\frac{6}{L^{2}}-\frac{1}{4m^2}
C_{\mu\nu\rho\sigma}C^{\mu\nu\rho\sigma}],
\end{eqnarray}
where $C_{\mu\nu\rho\sigma}$ is the Weyl tensor. In fact the action
(\ref{3'}) is the action of Einstein-Weyl gravity. The linearised
fluctuations around the $AdS_4$ vacuum in Einstein-Weyl gravity were
investigated in \cite{14}. This theory admits $dS_4$ space-time as
the vacuum solution of the field equations as well. Here we want to
solve the equation of motion in $4-$dimensional de Sitter background
with curvature  $- \frac{12}{L^{2}}$. The linearized equation of
motion in the following transverse traceless gauge

\begin{eqnarray}\label{2}
g^{\mu \nu}h_{\mu\nu}=0, ~~~~~~~~~~\nabla^\mu h_{\mu\nu}=0.
\end{eqnarray}
becomes
\begin{eqnarray}\label{3}
\left(\square+\frac{4}{L^2}-2m^{2}\right)\left(\square+\frac{2}{L^2}\right)h_{\mu\nu}=0.
\end{eqnarray}
The mass of massive mode is
\begin{eqnarray}\label{4'}
M^2=2(m^2-\frac{1}{L^2}).
\end{eqnarray}
In the critical case, $m^2=\frac{1}{L^2}$, in this limit the massive
mode becomes massless, and the two quadratic differential operators
degenerate.
 So, in the critical point,
we have
\begin{eqnarray}\label{1}
\left(\square+\frac{2}{L^2}\right)^{2}h_{\mu\nu}=0.
\end{eqnarray}
 We can write down $\square$ according to the Killing vectors of de Sitter spacetime and classify the answers of (\ref{1})
  according to the eigenstates of them. We can imagine de Sitter space as a $4-$dimensional  hypersurface in a $5-$dimensional
   spacetime:
\begin{eqnarray}\label{4}
ds^2=dx^2+dy^2+dz^2+du^2-dt^2, \nonumber
\\x^2+y^2+z^2+u^2-t^2=L^2.~~~~
\end{eqnarray}
The isometry group of the hypersurface is SO(1,4). Indeed, there are
6 Killings for rotations and 4 Killings for boosts:
\begin{eqnarray}\label{5}
&&L_{XY}=Y\partial{X}-X\partial{Y},\\
\nonumber &&L_{Xt}=t\partial{X}+X\partial{t},
\\
\nonumber &&X,Y={{x,y,z,u}}.
\end{eqnarray}
We can easily transform coordinates and work with this metric:
\begin{eqnarray}\label{6}
 ds^{2}=L^2(d\tau^{2}\cosh(\rho)^{2}+d\rho^{2}+\sinh(\rho)^{2}d\Omega^{2}),
\end{eqnarray}
where $d\Omega^{2}$ is the metric on $S^2$ with unit radius.  So two
of the Killing vectors are now the generators of transformation
along $\tau$ and $\rho$ axes:
\begin{eqnarray}\label{7}
H_1=-L_{ut}=-\partial_{\tau},~~~~~H_2=iL_{xy}=-i\partial_\varphi,
\end{eqnarray}
Indeed these are Cartan subalgebra of $so(1,4)$. Now, by choosing
the proper linear combination of other Killing vectors, they can
satisfy ladder commutation relations:
\begin{eqnarray}\label{8}
\nonumber E^{\alpha _{1}}&=&\frac{1}{2}e^{\left(\tau +i\varphi
\right)}[-\sin \left(\theta \right)\tanh \left(\rho \right) \partial
_{\tau }+\sin \left(\theta \right)\partial _{\rho }+\cos
\left(\theta \right)\coth \left(\rho \right)\partial _{\theta }
\\ \nonumber
&& +i\csc \left(\theta \right)\coth \left(\rho \right)\partial
_{\varphi }],
\\ \nonumber
E^{\alpha
_{2}}&=&e^{i\phi}[(i\partial_\theta-cot(\theta)\partial_\varphi],
\\ \nonumber
E^{\alpha _{3}}&=&\frac{1}{2}e^{\left(\tau -i\varphi \right)}[-\sin
\left(\theta \right)\tanh \left(\rho \right) \partial _{\tau }+\sin
\left(\theta \right)\partial _{\rho }+\cos \left(\theta \right)\coth
\left(\rho \right)\partial _{\theta }
\\ \nonumber
&&-i\csc \left(\theta \right)\coth \left(\rho \right)\partial
_{\varphi }],
\\ \nonumber
E^{\alpha_{4}}&=&e^\tau[\cos(\theta)\tanh(\rho)\partial_\tau
-\cos(\theta)\partial_\rho+\coth(\rho)\sin(\theta)\partial_\theta],
\\ \\ \nonumber
E^{-\alpha _{1}}&=&\frac{1}{2}e^{-\left(\tau +i\varphi \right)}[\sin
\left(\theta \right)\tanh \left(\rho \right) \partial _{\tau }+\sin
\left(\theta \right)\partial _{\rho }+\cos \left(\theta \right)\coth
\left(\rho \right)\partial _{\theta }
\\ \nonumber
&&-i\csc \left(\theta \right)\coth \left(\rho \right)\partial
_{\varphi }],
\\ \nonumber
E^{-\alpha
_{2}}&=&e^{-i\varphi}[(i\partial_\theta+cot(\theta)\partial_\varphi],
\\ \nonumber
E^{-\alpha _{3}}&=&\frac{1}{2}e^{-\left(\tau -i\varphi \right)}[\sin
\left(\theta \right)\tanh \left(\rho \right) \partial _{\tau }+\sin
\left(\theta \right)\partial _{\rho }+\cos \left(\theta \right)\coth
\left(\rho \right)\partial _{\theta }
\\ \nonumber
&&+i\csc \left(\theta \right)\coth \left(\rho \right)\partial
_{\varphi }],
\\ \nonumber
E^{-\alpha_{4}}&=&e^{-\tau}[-\cos(\theta)\tanh(\rho)\partial_\tau
-\cos(\theta)\partial_\rho+\coth(\rho)\sin(\theta)\partial_\theta],
\end{eqnarray}
Basically, these Killing vectors, except an $i$ factor for $\tau$,
are the same as in \cite{18}. Indeed, the transformation
$\tau\rightarrow i\tau$, convert (\ref{6}) to the AdS metric chosen
in \cite{18}. By the above selections for killing vectors, the
commutation relations are the same noted in \cite{18}
\begin{eqnarray}\label{9}
&&[H_i,E^{-\alpha_x}]=-\alpha^i_x
E^{-\alpha_x},~~~~~[H_i,E^{\alpha_x}]=\alpha^i_x E^{\alpha_x}
\nonumber \\ &&[H_1,H_2]=0,~~~~~~[E^{\alpha_x},E^{-\alpha_x}]=
\frac{2}{|\alpha_x|^{2}}\alpha_x.H,
\end{eqnarray}
which the second one demonstrate $E^{\alpha_x}$ generators as ladder
operators with the following values for $\alpha$ vectors:
\begin{eqnarray}\label{10}
\overrightarrow{\alpha_1}=(-1,1),~~~~\overrightarrow{\alpha_2}=(0,1),~~~~
\overrightarrow{\alpha_3}=(-1,-1),~~~~\overrightarrow{\alpha_4}=(-1,0).
\end{eqnarray}
Now we can construct Casimir operator by adding proper combinations
of squared Killing vectors and then write the equation of motion
(\ref{3}) according to it
\begin{eqnarray}\label{11}
\ell=H_1 H_1+H_2
H_2+\sum_{x=1}^{4}\frac{|\alpha_x|^2}{2}(E^{\alpha_x}E^{-\alpha_x}+E^{-\alpha_x}E^{\alpha_x}).
\end{eqnarray}
by acting $\ell$ on $h_{\mu\nu}$ we would obtain:
\begin{eqnarray}\label{12}
L^2\square h_{\mu \nu}=(\ell-8)h_{\mu \nu}.
\end{eqnarray}
So, equation (\ref{3}) converts to:
\begin{eqnarray}\label{13}
\left(\ell-6-m^{2}L^2\right)\left(\ell-6\right)h_{\mu\nu}=0.
\end{eqnarray}
Now we can solve this equation for those metric perturbations which
are the eigenstates of the Cartan subalgebra of $so(1,4)$:
\begin{eqnarray}\label{14}
H_1 \psi_{\mu \nu}=E_0\psi_{\mu \nu},~~~~~~~~~~H_2 \psi_{\mu
\nu}=s\psi_{\mu \nu}.
\end{eqnarray}
For each root vector, we can construct the $su(2)$ subalgebra:
\begin{eqnarray}\label{15}
E^\pm=|\alpha_x|^{-1}E^{\pm\alpha_x},~~~~~~~~E^3=|\alpha_x|^{-2}\alpha.H,
\end{eqnarray}
then for the highest weight state, we have:
\begin{eqnarray}\label{16}
E^{\alpha_1}\psi_{\mu \nu}=0,~~~~~E^{\alpha_3}\psi_{\mu
\nu}=0,~~~~~E^{\alpha_4}\psi_{\mu \nu}=0.
\end{eqnarray}

 so $s$ and $E_0$ can be
interpreted as the eigenvalues of the energy and angular momentum of
states. We can choose the below ansatz for $\psi_{\mu \nu}$
\cite{18,19}:
\begin{eqnarray}\label{15}
&&\psi_{\mu \nu}=f(\tau,\rho, \theta, \phi) \Omega_{\mu \nu},
\\ \nonumber
&&\Omega_{\mu \nu}=\begin{pmatrix}
                 -1 & ia & ib &-1\\
                 ia&-a^2&-ab & ia\\
                 ib&-ba &-b^2& ib\\
                 -1& ia & ib & 1 \\
                 \end{pmatrix},
\end{eqnarray}
where $a$ and $b$ are only functions of $\rho$ and $\theta$. This
ansatz ensure the condition $\nabla^\mu h_{\mu\nu}=0$. $\psi_{\mu
\nu}$ is traceless, so:
\begin{eqnarray}\label{16}
a=\frac{1}{\sinh(\rho)\cosh(\rho)}, ~~~~~~~b=\cot(\theta).
\end{eqnarray}
If we assume $f(\tau,\rho,\theta,\phi)=T(\tau)R(\rho)\Theta(\theta)
\Phi(\phi)$, relations (\ref{14}) and (\ref{16}) imply that:
\begin{eqnarray}\label{18}
f(\tau,\rho,\theta,\phi)=e^{E_0\tau}e^{is\phi} \sin(\theta)^s
\sinh(\rho)^s \cosh(\rho)^{-E_0}
\end{eqnarray}
we can obtain lower states of energy by apply of $E^{-\alpha_4}$ and
lower states of angular momentum by $E^{\alpha_2}$.
\\
Now, we can consider the effect of the Casimir operator on
$\psi_{\mu \nu}$:
\begin{eqnarray}\label{20}
\ell\psi_{\mu \nu}=(E_0(E_0-3)+s(s+1))\psi_{\mu \nu}.
\end{eqnarray}
So, (\ref{13}) for $\psi_{\mu \nu}$ leads to:
\begin{eqnarray}\label{21}
&&(E_0(E_0-3)+s(s+1)-6)=0,
\nonumber \\
&&or
\nonumber \\
&&(E_0(E_0-3)+s(s+1)-6-m^2L^{2}=0.
\end{eqnarray}
We can obtain two $E_0$ from these equations which lead to two
answers for (\ref{18}). Any linear combination of these two modes
still is an answer. In the massless limit, there would be a
degeneracy and both equations imply $s\leq2$. We can find a log mode
by the below combination \cite{19}:
\begin{eqnarray}\label{22}
\psi_{\mu \nu}^{log}=(\frac{3}{m})^2(\psi_{\mu
\nu}^{E_0^{(1)},s}-\psi_{\mu \nu}^{E_0^{(2)},s}).
\end{eqnarray}
For $s=2$, in the limit of $m\rightarrow0$ we would have:
\begin{eqnarray}\label{23}
\psi_{\mu \nu}^{log}=\pm(\tau-\ln(\cosh(\rho))\psi_{\mu
\nu}^{E^{(1)}_0},
\end{eqnarray}
the positive sign is for $E^{(1)}_0=3$ and the minus sign is for
$E^{(1)}_0=0$. These log modes correspond to eigenstates of $H_2$

\begin{eqnarray}\label{23'}
H_2\psi_{\mu \nu}^{log}=s\psi_{\mu \nu}^{log},
\end{eqnarray}

 The effect of $H_1$ on this state is:
\begin{eqnarray}\label{24}
H_1\psi_{\mu \nu}^{log}=\pm\psi_{\mu
\nu}^{E^{(1)}_0}+E^{(1)}_0\psi_{\mu \nu}^{log},
\end{eqnarray}
so $H_1$ is not diagonal and the log modes in de Sitter space-time
obey similar properties as the ones in AdS-space-time \cite{18}.
Since the generator $H_1$ can be identified with the conformal
energy operator in the dual field theory, then Eq.(\ref{24})
indicates that this conformal energy is not diagonalizable, which is
the defining property of a logarithmic CFT. Therefore this result
would be a strong indication that critical gravity theories in de
Sitter space-times could lead to a de Sitter/log CFT correspondence.

\section{Conclusion}
In the present paper we have considered the critical gravity in
$4-$dimensional de Sitter space-time. The Lagrangian consists of the
Einstein-Hilbert term with a cosmological constant $\Lambda$ and an
additional higher-order term proportional to the square of the Weyl
tensor, with a coupling constant \cite{14}. It was shown that there
is a critical relation between coupling constant and $\Lambda$ for
which the generically-present massive spin-2 modes disappear, and
are instead replaced by modes with a logarithmic fall-off
\cite{16,18}. These log modes are ghostlike, but since they fall off
more slowly than do the massless spin-2 modes, they can be truncated
out by imposing an appropriate AdS boundary condition \cite{20,21}.
The resulting theory is then somewhat trivial, in the sense that the
remaining massless graviton has zero on-shell energy. Furthermore,
the mass and the entropy of standard Schwarzschild-AdS black holes
are both zero in the critical theory. Now may be one ask why the
study of these models are important, and why the existence of these
logarithmic solutions is interesting? In answer to these question we
can say that these models could provide gravity descriptions for
 logarithmic CFTs. Logarithmic CFT arise in various contexts in condensed matter
 physics \cite{22}. A defning feature of logarithmic CFTs is that the Hamiltonian does
not diagonalise, but rather contains Jordan cells of rank two or
higher. In the AdS-case, where the generator $H_1$ can be identified
with the conformal energy operator in the dual field theory,
Bergshoeff et. al\cite{18} have shown that  this Hamiltonian is not
diagonalizable, which is the defining property of a logarithmic CFT.
In this paper we extended this work to the de Sitter space. We have
obtained the logarithmic mode for $s=2$ case by Eq.(\ref{23}). Then
we have shown that these logarithmic modes in de Sitter space-time
obey similar properties as the ones in AdS-space-time. Namely
 the properties of Eq. (64) and especially Eq. (65) in
paper \cite{18} still hold. These equations would be a strong
indication that critical gravity theories in de Sitter space-times
could also lead to a de Sitter/log CFT correspondence. Although de
Sitter/CFT and the subject is certainly not as well established as
the usual AdS/CFT correspondence, but a de Sitter/log CFT conjecture
could certainly be interesting to speculate about (since log CFT's
have found applications in condensed matter physics and possible
gravity duals might thus be interesting).

\section{Acknowledgements}
We thank Prof. Andrew Strominger and Dr. Jan Rosseel for helpful
discussions and correspondence.

\end{document}